\documentclass{aa}
\usepackage{graphicx,amssymb}
\usepackage{txfonts}
\begin{document}
\title{The afterglow and the host galaxy of 
GRB 011211\thanks{Based on observations 
made with the Nordic
Optical Telescope, operated on the island of La Palma jointly by
Denmark, Finland, Iceland, Norway, and 
Sweden.}$^{,}$\thanks{Based on observations made with 
ESO Telescopes at the Paranal Observatory by GRACE under programme 
ID \mbox{69.D-0701}.}$^{,}$\thanks{Based on observations made with the 
NASA/ESA Hubble Space Telescope, 
obtained from the Data Archive at the Space Telescope Science 
Institute, which is operated by the Association of Universities 
for Research in Astronomy, Inc., under NASA contract NAS 5-26555. 
These observations are associated with program \#8867.}}

\author{P. Jakobsson
       \inst{1,2} \and
       J. Hjorth
       \inst{1} \and
       J. P. U. Fynbo
       \inst{1,3} \and
       J. Gorosabel
       \inst{4,5} \and
       K. Pedersen
       \inst{1} \and 
       I. Burud 
       \inst{5} \and
       A. Levan
       \inst{2,5} \and
       C.~Kouveliotou
       \inst{6} \and
       N. Tanvir
       \inst{7} \and
       A. Fruchter
       \inst{5} \and
       J. Rhoads
       \inst{5} \and
       T. Grav
       \inst{8} \and
       M. W. Hansen
       \inst{8} \and
       R. Michelsen
       \inst{1} \and
       M.~I.~Andersen
       \inst{9} \and
       B. L. Jensen
       \inst{1} \and
       H.~Pedersen
       \inst{1} \and
       B. Thomsen
       \inst{3} \and
       M. Weidinger 
       \inst{3} \and
       S. G. Bhargavi
       \inst{10} \and
       R.~Cowsik
       \inst{10} \and
       S. B. Pandey
       \inst{11} 
}

\offprints{P. Jakobsson, \\ \email{pallja@astro.ku.dk}}
   
\institute{Astronomical Observatory, University of Copenhagen,
     Juliane Maries Vej 30, 2100 Copenhagen \O, Denmark
\and
     Department of Physics and Astronomy, University of Leicester, 
     University Road, Leicester, LE1 7RH, UK
\and
     Department of Physics and Astronomy, 
     University of Aarhus, Ny Munkegade, 8000 \AA rhus C, Denmark
\and
     IAA-CSIC, P.O. Box 03004, E-18080 Granada, Spain
%     Danish Space Research Institute, Juliane Maries Vej 30, DK-2100
%     Copenhagen \O, Denmark 
\and
     Space Telescope Science Institute, 3700 San Martin Drive,
     Baltimore, MD 21218, USA
\and
     NSSTC, SD-50, 320 Sparkman Drive, Huntsville, Alabama 35805, USA
\and
     Department of Physical Sciences, University of Hertfordshire,
     College Lane, Hatfield, Herts AL10 9AB, UK 
\and
     Institute of Theoretical Astrophysics, University of Oslo, PB
     1029 Blindern, 05315 Oslo, Norway 
\and
     Astrophysikalisches Institut Potsdam, An der Sternwarte 16,
     D-14482 Potsdam, Germany
\and
     Indian Institute of Astrophysics, Sarjapur Road, 
     Bangalore 560 034, India
\and
     State Observatory, Manora Peak, Nainital 263 129, India
}
   \date{Received 2 November 2002 / Accepted 7 July 2003}

\abstract{We present optical, near-infrared, and X-ray observations of the 
optical afterglow (OA) of the X-ray rich, long-duration gamma-ray 
burst GRB~011211. Hubble Space Telescope (HST) data obtained 14, 26,
32, and 59 days after the burst, show the host galaxy to have a
morphology that is fairly typical of blue galaxies at high redshift.
We measure its magnitude to be $R = 24.95 \pm 0.11$. We detect a 
break in the OA $R$-band light curve which is naturally accounted 
for by a collimated outflow geometry. By fitting a broken power-law 
to the data we find a best fit with a break $1.56 \pm 0.02$~days
after the burst, a pre-break slope of $\alpha_1 = -0.95 \pm 0.02$, 
and a post-break slope of $\alpha_2 = -2.11 \pm 0.07$. 
The UV-optical spectral energy distribution (SED) around 14 hours
after the burst is best fit with a power-law  
with index $\beta = -0.56 \pm 0.19$ reddened by an SMC-like 
extinction law with a modest $A_V = 0.08 \pm 0.08$ mag. 
By comparison, from the XMM-Newton X-ray data at around the same time,
we find a decay index of 
$\alpha_{\mathrm{X}} = -1.62 \pm 0.36$ and a spectral index of 
$\beta_\mathrm{X} = -1.21^{+0.10}_{-0.15}$. Interpolating between the 
UV-optical and X-ray implies that the cooling
frequency is located close to $\sim$10$^{16}$~Hz in the 
observer frame at the time of the
observations. We argue, using the various temporal and spectral
indices above, that the most likely afterglow model is that of a
jet expanding into an external environment that has a constant 
mean density rather than a wind-fed density structure. We estimate 
the electron energy index for this burst to be 
$p \sim2.3$.
}

\titlerunning{The afterglow and the host galaxy of GRB 011211}

\authorrunning{P. Jakobsson et al.}

\maketitle
%----------------------INTRODUCTION-------------------------------
\section{Introduction}
A deceleration of a relativistic fireball in the surrounding
environment 
is now widely believed to cause the afterglow emission of GRBs 
(see Piran \cite{piran} for a review). The external
medium could be either a precursor wind from the GRB progenitor
(Chevalier \& Li \cite{che}) or the interstellar medium (ISM) of the
host galaxy (Waxman \cite{wax}). The interaction between the 
fireball and this ambient medium produces a shock that 
accelerates electrons and gives
them a power-law distribution of ultra-relativistic energies, $N(\gamma) 
\propto \gamma^{-p}$, where $p$ is the electron energy index. This
leads to the production of synchrotron emission where the flux of the
afterglow can be described by a power-law decline in time and frequency, 
$F \propto t^{\alpha} \nu^{\beta}$ (Sari, Piran \& Narayan \cite{sari98}). 
The decay rate, $\alpha$, depends on the
nature of the fireball and also on the density structure of the
ambient medium. 
Light curves from observed afterglows
typically have an initial decay index of 
$\alpha_1 \sim -1$, which often steepens 1--3 days after the GRB to 
$\alpha_2 \sim -2$ or even steeper (e.g. Fig.~4 in Andersen et al. 
\cite{andersen}). 
\par
The GRB~011211 $R$-band light curve, presented by Holland 
et al. (\cite{holland2}, hereafter H02), showed the OA decaying as a
power-law with a slope of $\alpha = -0.83 \pm 0.04$ for the first
$\sim$2~days after the burst at which time there was evidence for a
break. Reeves et al. (\cite{reeves}, hereafter R02) found that the 
X-rays emitted in the wake of GRB~011211 originated in an extremely 
hot gas outflowing from the GRB progenitor at $\sim$0.1$c$, and that 
this gas was highly enriched with the by-products of a supernova explosion. 
\par
In this paper we present photometry of the OA of GRB~011211, 
taken between $\sim$0.6 and $\sim$30 days after the
burst occurred. We explore the properties of the X-ray light curve,
observed between $\sim$0.5 and $\sim$0.85 days from the onset of the
burst. We model the afterglow data and conclude that the most likely 
model is a jet expanding into an external environment with a 
constant mean density. We also analyse HST/STIS images in order 
to derive the photometric properties of the host galaxy. 
\par
The organization of this paper is as follows. The optical and
near-infrared (NIR) observations are presented in
Sect.~\ref{obs.sec}. In Sect.~\ref{hst.sec} we analyze HST/STIS images
of the OA and the host galaxy. In Sects.~\ref{optical.sec} and \ref{xray.sec}
we investigate the optical and X-ray light curves. The spectral 
energy distribution (SED) of the
afterglow along with the derived extinction is discussed in
Sect.~\ref{sed.sec}. We use the derived properties of the OA in 
Sect.~\ref{models.sec} to compare our results with afterglow
models. Finally, Sect.~\ref{dis.sec} summarises the main results.
Throughout this paper, we adopt a Hubble constant of \mbox{$H_0 = 
65 \ \textrm{km s}^{-1} \ \textrm{Mpc}^{-1}$} and assume $\Omega_{m}
= 0.3$ and \mbox{$\Omega_{\Lambda} = 0.7$}.
%-----------------------OBSERVATIONS-------------------------------
\section{Ground-based observations}
\label{obs.sec}
GRB 011211 was detected by the Italian/Dutch satellite BeppoSAX on
2001 Dec
11.7982 UT. A $5\arcmin$ radius error circle was circulated via the
GRB Coordinate Network
(GCN)\footnote{\texttt{http://gcn.gsfc.nasa.gov/gcn/}} 5.4 hours after
the burst (Gandolfi \cite{gandolfia}). Less than 2 hours later the
error radius was refined to only $2\arcmin$ (Gandolfi
\cite{gandolfib}). This X-ray rich GRB had a duration of 
approximately 270~s, making it one of the longest duration bursts 
observed by BeppoSAX and thus placing it firmly in 
the ``long-duration'' burst category.
\par
%-------------------FIGURE-----------------------------------------
   \begin{figure}
   \centering
   \resizebox{\hsize}{!}{\fbox{\includegraphics{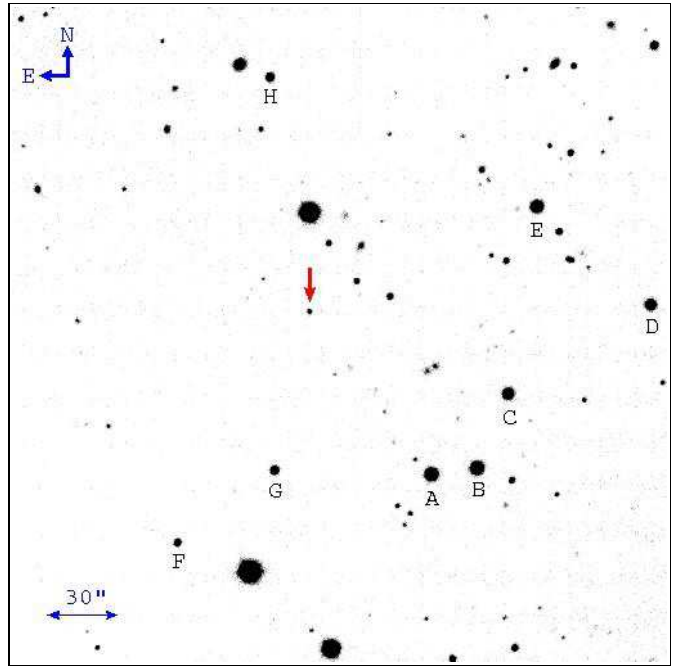}}}
      \caption{An $R$-band image of the GRB~011211 optical
   afterglow. The position of the optical afterglow is marked 
   with an arrow. Also
   marked are the eight secondary calibration stars A--H used to
   transform the relative photometry onto the standard system. The
   photometric properties of these stars are given in Table~\ref{hend.tab}.}
         \label{ot.fig}
   \end{figure}
%----------------------END FIGURE---------------------------------
The OA was discovered at the Nordic Optical Telescope (NOT) (Grav et
al. \cite{grav}; Jakobsson 
et al. \cite{palli}, hereafter J03) on 2001 Dec 12.2 UT, 9.6 hours 
after the burst. A NOT image of the OA and its surrounding field 
is displayed in Fig.~\ref{ot.fig}. The OA was monitored 
during the following week with
the NOT and with the Danish \mbox{1.54-m} telescope at La Silla (Jensen et
al. \cite{brian2}). The decline
of the OA was also observed at several other telescopes (Soszynski et
al. \cite{sos}; Bhargavi \& Cowsik \cite{bhargavi};
Holland et al. \cite{holland}). 
In India the follow-up observations of the OA
were carried out between 2001 Dec 12--14, using the \mbox{2.34-m} 
Vainu Bappu Telescope (VBT) as well
as the \mbox{1-m} Zeiss telescope of the Vainu Bappu Observatory (VBO), and 
on 2001 Dec 12, using the \mbox{1-m} Sampurnanda telescope at 
the State Observatory (SO). In addition, the 3.58-m New Technology 
Telescope (NTT) at La Silla was used to obtain NIR images 
of the OA shortly after the burst, while $R$-band images $\sim$5.5 and 
$\sim$9.5 days after the burst were obtained at the Melipal and Yepun of
the \mbox{8.2-m} Very Large Telescope (VLT) as part of the 
GRACE\footnote{\underline{GR}B \underline{A}fterglow 
\underline{C}ollaboration at \underline{E}SO.} program. 
The journal of our observations is listed in Table~\ref{phot.tab}.
\par
The redshift of the burst was measured
to be $z = 2.140$ via absorption lines in an optical spectrum taken
at the Yepun of the VLT (Fruchter et
al. \cite{fruchter}). This redshift was confirmed with spectra
obtained at the Magellan \mbox{6.5-m} Walter Baade telescope (Gladders et
al. \cite{gladders}; H02) and with the 
\mbox{3.5-m} Telescopio Nazionale Galileo (Fiore et al. \cite{fiore}).
%---------------------------------TABLE----------------------------
\begin{table}[!t]
\caption[]{\normalsize The journal of the GRB~011211 observations and
the results of the photometry. Upper limits are $2 \sigma$ in a
circular aperture with radius $2\arcsec$. No corrections for
extinction have been applied to the photometry in the table. The flux
from the host galaxy has not been subtracted.
Notes: $^a$ for the last two HST data 
points the date is 2002 January; $^b$ OA magnitude.}
     \label{phot.tab}
\centering
%\tiny
\scriptsize 
%\footnotesize
\setlength{\arrayrulewidth}{0.8pt}   % Default is 0.4pt
\begin{tabular}{lccccc}
\hline
\hline
\vspace{-2 mm} \\
Date (UT) & Obs. & Magnitude & Seeing & Exp. time\\
of 2001 Dec$^a$ & & & (arcsec) & (s)\\
\vspace{-2 mm} \\
\hline
\vspace{-2 mm} \\
\hspace{-3 mm}
\emph{U-band:}    &     &                  &      & \\
13.2467 & NOT  & $21.5  \pm 0.3$  & 1.20 & $3 \times
600$ \\
\hspace{-3 mm}
\emph{B-band:}   &      &                    &      & \\
%12.2816 & NOT   & $21.229 \pm 0.047$ & 1.10 & 300 \\
%\textbf{12.3539} & 1.54-m & $21.262 \pm 0.089$ & 1.35 & 200 \\
12.9819 & VBT   & $21.914 \pm 0.108$ & 2.5  & 1800 \\
13.2280 & NOT   & $22.241 \pm 0.076$ & 1.55 & $3 \times 300$ \\
13.3382 & 1.54-m & $22.287 \pm 0.068$ & 1.00 & $3 \times 600$ \\
13.9550 & VBT   & $23.06 \pm 0.27$   & 3.00 & $2 \times 1200$ \\
\hspace{-3 mm}
\emph{V-band:}    &            &                    &      & \\
%12.2907 & NOT   & $20.761 \pm 0.080$ & 1.00 & 300 \\
%\textbf{12.3586} & 1.54-m & $20.987 \pm 0.042$ & 1.20 & 300 \\
12.9403 & VBT   & $21.522 \pm 0.077$ & 3.00 & 1200 \\
12.9569 & VBT   & $21.575 \pm 0.073$ & 2.44 & 1200 \\
13.2189 & NOT   & $21.478 \pm 0.111$ & 1.55 & 300 \\
13.3044 & 1.54-m & $21.757 \pm 0.089$ & 1.00 & $3 \times 300$ \\
13.9180 & VBT   & $22.82 \pm 0.22$   & 2.80 & $2 \times 900$ \\
20.3087 & 1.54-m & $25.00 \pm 0.22$   & 1.35 & $7 \times 1200$ \\
\hspace{-3 mm}
\emph{R-band:}    &      &                    &      & \\
%12.2181 & NOT   & $20.012 \pm 0.033$ & 1.70 & 300 \\
%12.2227 & NOT   & $19.877 \pm 0.029$ & 1.40 & 300 \\
%12.2272 & NOT   & $19.918 \pm 0.022$ & 1.30 & 300 \\
%12.2831 & 1.54-m & $20.368 \pm 0.040$ & 1.50 & 600 \\
%12.2912 & 1.54-m & $20.298 \pm 0.033$ & 1.45 & 600 \\
%12.2994 & 1.54-m & $20.230 \pm 0.038$ & 1.40 & 600 \\
%12.3075 & 1.54-m & $20.285 \pm 0.042$ & 1.35 & 600 \\
%12.3157 & 1.54-m & $20.274 \pm 0.051$ & 1.30 & 600 \\
%12.3238 & 1.54-m & $20.330 \pm 0.046$ & 1.10 & 600 \\
%12.3633 & 1.54-m & $20.539 \pm 0.063$ & 1.10 & 300 \\
%\textbf{12.3681} & 1.54-m & $20.616 \pm 0.100$ & 1.00 & 300 \\
12.8722 & VBT   & $20.91  \pm 0.13$  & 2.20 & 600 \\
12.8872 & VBT   & $21.022 \pm 0.089$ & 2.20 & 900 \\
12.9007 & VBT   & $20.732 \pm 0.081$ & 2.10 & 600 \\
12.9097 & VBT   & $20.952 \pm 0.092$ & 2.10 & 600 \\
12.9188 & VBT   & $20.959 \pm 0.094$ & 2.10 & 600 \\
12.9271 & VBT   & $21.061 \pm 0.129$ & 2.10 & 600 \\
12.9399 & VBO & $21.204 \pm 0.168$ & 2.70 & 900 \\
12.9602 & SO    & $21.037 \pm 0.067$ & 3.40 & $3 \times 300$ \\
12.9753 & VBO & $21.203 \pm 0.163$ & 2.10 & 900 \\
12.9785 & SO    & $21.174 \pm 0.047$ & 3.40 & $2 \times 600$ \\
12.9945 & SO    & $21.134 \pm 0.063$ & 3.40 & $2 \times 600$ \\
12.9979 & VBT   & $21.100 \pm 0.110$ & 2.30 & 600 \\
13.0108 & SO    & $21.195 \pm 0.088$ & 3.40 & $2 \times 600$ \\
13.2096 & NOT   & $21.123 \pm 0.076$ & 1.10 & 300 \\
13.2843 & 1.54-m & $21.378 \pm 0.067$ & 1.20 & 600 \\
13.2924 & 1.54-m & $21.408 \pm 0.062$ & 0.95 & 600 \\
13.3550 & 1.54-m & $21.356 \pm 0.057$ & 0.85 & 600 \\
13.3631 & 1.54-m & $21.495 \pm 0.086$ & 0.95 & 600 \\
15.3438 & 1.54-m & $>23.0$            & 1.15 & $1200+900$ \\
16.3071 & 1.54-m & $23.32  \pm 0.25$  & 1.30 & $7 \times 1200$ \\
17.3031 & 1.54-m & $23.69  \pm 0.25$  & 1.80 & $6 \times 1200$ \\
17.3075 & VLT   & $24.1   \pm 0.2$   & 1.30 & 1800 \\
18.3044 & 1.54-m & $24.22  \pm 0.40$  & 1.10 & $6 \times 1200$ \\
%19.2947 & 1.54-m & $24.89  \pm 0.38$  & 1.20 & $9 \times 1200$ \\
21.3080 & VLT   & $25.27   \pm 0.16^b$   & 0.70 & 3600 \\
25.8190 & HST   & $26.71 \pm 0.16^b$   & ---  & 5193 \\
6.7720  & HST   & $27.45 \pm 0.21^b$   & ---  & 4785 \\
12.712  & HST   & $28.40 \pm 0.48^b$  & ---  & 4785 \\
\hspace{-3 mm}
\emph{I-band:}    &      &                    &      & \\
%12.2861 & NOT   & $19.965 \pm 0.110$ & 0.90 & 300 \\
%12.3350 & 1.54-m & $19.951 \pm 0.054$ & 1.30 & 300 \\
%12.3397 & 1.54-m & $19.981 \pm 0.071$ & 1.80 & 300 \\
%\textbf{12.3491} & 1.54-m & $20.064 \pm 0.061$ & 0.40 & 300 \\
%\textbf{12.3398} & 1.54-m & $20.011 \pm 0.087$ & 1.20 & $5 \times 300$ \\
12.9879 & VBO & $20.92 \pm 0.21$  & 2.60 & 900 \\
13.2143 & NOT   & $20.99  \pm 0.23$  & 1.00 & 300 \\
13.3237 & 1.54-m & $20.86  \pm 0.15$  & 0.90 & $3 \times 300$ \\
15.3069 & 1.54-m & $22.75  \pm 0.25$  & 1.15 & $4 \times 900$ \\
20.2350 & 1.54-m & $>23.2$            & 1.25 & $3 \times 900+500$ \\
\hspace{-3 mm}
\emph{J-band:}    &      &                    &      & \\
12.3616 & NTT   & $19.34  \pm 0.05$  & 1.05 & 900 \\
14.3470 & NTT   & $21.04  \pm 0.05$  & 0.55 & 2700 \\
\hspace{-3 mm}
\emph{K-band:}    &      &                    &      & \\
12.3736 & NTT   & $18.02  \pm 0.07$  & 0.90 & 840 \\
\vspace{-2 mm} \\
\hline
\end{tabular}
\end{table}
%---------------------------END TABLE--------------------------------
%---------------------The Host Galaxy---------------------------
\section{Observations from space}
\label{hst.sec}
In an XMM-Newton follow-up observation (0.51--0.84 days after the GRB
trigger) a source in the BeppoSAX \mbox{error} box was detected
(Santos-Lleo et al. \cite{leo}). Subsequent analysis showed this source
to be fading with a decay index $\alpha_{\mathrm{X}} = -1.7\pm 0.2$ in the
0.2--10 keV band (R02).
\par
\begin{figure}
   \centering
   \resizebox{\hsize}{!}{\includegraphics{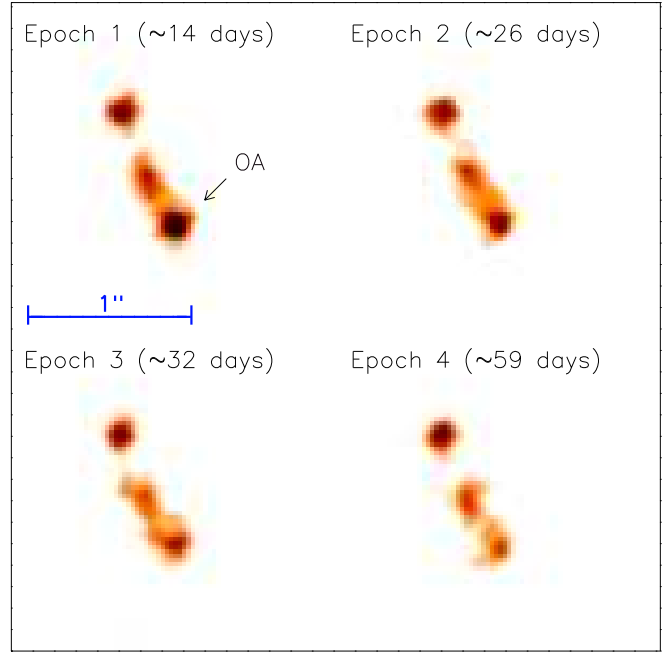}}
      \caption{HST images centered on the host galaxy at four
   different epochs showing the decline of the optical afterglow. 
   North is up and east is to the left. The numbers in the 
   parentheses indicate the time after the burst.}
         \label{decline.fig}
   \end{figure}

We have reduced the publicly available HST 
data\footnote{HST proposal 8867 (S. Kulkarni).} 
taken $\sim$14 (epoch 1), $\sim$26 (epoch 2), $\sim$32 (epoch 3) 
and $\sim$59 (epoch 4) days after 
the burst (Fig.~\ref{decline.fig}). The dithered data 
were pre-processed using
the "on the fly" calibration from the HST 
archive\footnote{\texttt{http://archive.stsci.edu}} and were drizzled onto
a final output grid with pixels half the size of the original input
pixels using a value for {\tt pixfrac} of 0.7 
(Fruchter \& Hook \cite{fruchter02}). In order to convert the
50CCD (clear, hereafter referred to as CL) broadband STIS 
magnitude to an $R$-band magnitude
we assumed a power-law spectrum of $\beta=-0.56 \pm 0.19$ as derived
in Sect.~\ref{sed.sec}. We note that the magnitude 
errors quoted include a term due to this
colour correction since the late time colour is poorly constrained.
This colour term would not normally be expected to change
substantially for a GRB afterglow, however an underlying supernova can
lead to significant reddening of the normally blue afterglow
spectrum. In practice this effect will be of the order of 0.1 mags in
changing the power-law from that observed to $\sim$$\nu^{-3}$, and
hence our corrections to an $R$-band magnitude allow for this
possibility. In order to calculate the magnitude of the OA we 
subtracted the epoch 4
image from the previous three images and then performed aperture
photometry. To avoid any contamination from
the OA in the final HST image, we subtracted from it the expected OA
magnitude (see Sect.~\ref{a.sec} and the dotted line in 
Fig.~\ref{curve.fig}). The host galaxy seems to
have a multi-component morphology, similar to the host of GRB~000926 
(Fynbo et al. \cite{johan}) and many other high-$z$ 
galaxies (e.g. Giavalisco et al. 1996; M\o ller et al. 2002). 
Recent Ly$\alpha$ observations (Fynbo et al. \cite{jp})
have revealed that all the components 
are related to the host, with most of
the Ly$\alpha$ emission emitted from the source north of the OA
position.
%-------------------FIGURE-----------------------------------------
%----------------------END FIGURE---------------------------------
\par
Using a large ($1\arcsec$) aperture we estimated the total magnitude of the
host galaxy complex in our broadband STIS image to be $\textrm{CL} \sim 
25.60 \pm 0.05$. 
In order to calculate the $R$-magnitude of the host we estimated a colour
correction based on the colours of other GRB host galaxies (e.g. 
Sokolov et al. \cite{sok}). Allowing for Galactic reddening
in the direction of GRB~011211, $E(B-V)=0.045$ from
Schlegel, Finkbeiner \& Davis (\cite{schlegel}), the range of 
reasonable colour corrections ($R-V$) for this
host lies between $-0.75$ and $-0.55$. We adopted the
centre of this range ($-0.65$ magnitudes) as the correction for the host of
GRB~011211. This implies that the $R$-magnitude of the host is 
$24.95 \pm 0.11$. We note that this is a somewhat crude estimate 
since the exact colour profile of the host is unknown. The results of
the OA photometry are presented in Table~\ref{phot.tab}.
%------------------The optical light curve--------------------------
\section{The optical light curve}
\label{optical.sec}
%---------------------------------TABLE----------------------------
\begin{table}
\caption[]{$BVRI$ magnitudes (Henden \cite{henden2}) for the eight 
secondary calibration stars A--H (see Fig.~\ref{ot.fig}).}
     \label{hend.tab}
\centering
%\tiny
\scriptsize 
\setlength{\arrayrulewidth}{0.8pt}   % Default is 0.4pt
\begin{tabular}{lcccc}
\hline
\hline
\vspace{-2 mm} \\
Star & \hspace{-0.35cm} $B$ & \hspace{-0.3cm} $V$ 
& \hspace{-0.35cm} $R$ & \hspace{-0.3cm} $I$ \\
\vspace{-2 mm} \\
\hline
\vspace{-2 mm} \\
A & \hspace{-0.35cm} $17.292 \pm 0.011$ & \hspace{-0.35cm} $16.392 \pm
0.011$ & 
\hspace{-0.35cm} $15.838 \pm 0.014$ & \hspace{-0.35cm} $15.310 \pm 0.020$ \\
B & \hspace{-0.35cm} $17.507 \pm 0.008$ & \hspace{-0.35cm} $16.481 \pm 0.008$ &
\hspace{-0.35cm} $15.849 \pm 0.015$ & \hspace{-0.35cm} $15.303 \pm 0.048$ \\
C & \hspace{-0.35cm} $19.069 \pm 0.034$ & \hspace{-0.35cm} $17.520 \pm 0.034$ &
\hspace{-0.35cm} $16.441 \pm 0.038$ & \hspace{-0.35cm} $15.056 \pm 0.060$ \\ 
D & \hspace{-0.35cm} $17.824 \pm 0.015$ & \hspace{-0.35cm} $17.158 \pm 0.008$ &
\hspace{-0.35cm} $16.744 \pm 0.015$ & \hspace{-0.35cm} $16.401 \pm 0.019$ \\
E & \hspace{-0.35cm} $17.128 \pm 0.009$ & \hspace{-0.35cm} $16.479 \pm 0.008$ &
\hspace{-0.35cm} $16.096 \pm 0.020$ & \hspace{-0.35cm} $15.706 \pm 0.043$ \\
F & \hspace{-0.35cm} $20.874 \pm 0.129$ & \hspace{-0.35cm} $19.446 \pm 0.036$ &
\hspace{-0.35cm} $18.389 \pm 0.055$ & \hspace{-0.35cm} $17.186 \pm 0.061$ \\
G & \hspace{-0.35cm} $18.479 \pm 0.016$ & \hspace{-0.35cm} $18.002 \pm 0.011$ &
\hspace{-0.35cm} $17.705 \pm 0.021$ & \hspace{-0.35cm} $17.383 \pm 0.104$ \\
H & \hspace{-0.35cm} $18.768 \pm 0.060$ & \hspace{-0.35cm} $18.130 \pm 0.040$ &
\hspace{-0.35cm} $17.783 \pm 0.046$ & \hspace{-0.35cm} $17.304 \pm 0.183$ \\ 
\vspace{-2 mm} \\
\hline
\end{tabular}
\end{table}
%---------------------------END TABLE--------------------------------
\subsection{Construction of the R-band light curve}
\label{curve}
We measured the magnitude of the OA relative to 8 stars in the
field. The calibrated magnitudes of these stars are given in Henden
(\cite{henden}, \cite{henden2}) and shown 
in Table~\ref{hend.tab}. In some of our 
images we only used a subset of these 8
stars due to saturation. We used aperture photometry in a circular 
aperture with radius
$2\arcsec$ in order to fully include the emission from the host galaxy
reported by Burud et al. (\cite{burud}). Finally, we used our HST
results to subtract the host contribution from each data point.
\par
\subsection{Power-law fitting}
\label{a.sec}
\begin{figure}
   \centering
\resizebox{\hsize}{!}{\includegraphics{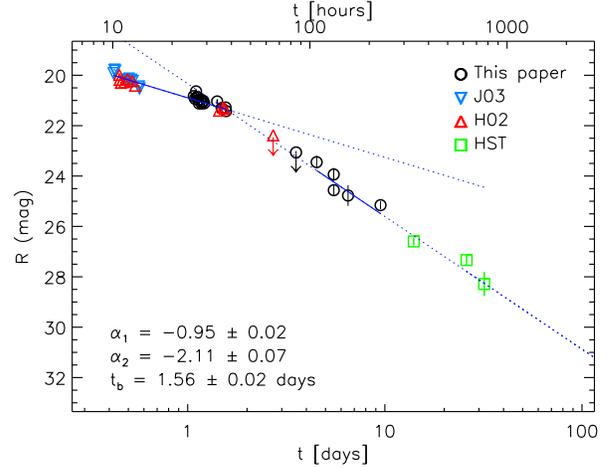}}
      \caption{The $R$-band light curve of the OA based on the
   measurements given in Table~\ref{phot.tab} (circles), the data
   presented in H02 (triangles) and J03 (upside down triangles), 
   and publicly available HST data 
   (squares). All data points have been corrected for Galactic
   reddening. The dotted lines are extrapolation of the straight 
   lines fit through the light curves.}
         \label{curve.fig}
   \end{figure}

Our $R$-band light curve is shown in Fig.~\ref{curve.fig}. It is
supplemented by early light curve data points from H02 and J03. 
Also plotted is a broken power-law fit to
the light curve prior to day 10. From the formal best fit we find that 
the initial light curve decay has a power-law index of
$\alpha_1 = -0.95 \pm 0.02$, while $\alpha_2 = -2.11 \pm 0.07$, with the 
break occurring at $t_{\mathrm{b}} = 1.56 \pm 0.02$~days 
($\chi^2_{43} = 7.82$, where $\chi^2_{\mathrm{dof}}= \chi^2/
\mathrm{degree~of~freedom}$, is the reduced $\chi^2$ of the fit). 
Neither the three HST points, nor the two upper limits were
included in this fit. In the case of the HST points, there is
a risk of contamination from a possible supernova (SN) bump, 
however, including them does not 
affect $\alpha_1$ or $t_{\mathrm{b}}$, and $\alpha_2$ 
only changes to $-2.13 \pm 0.04$.
\par
The broken power-law fits are formally strongly rejected by the data
due to the wiggles in the early light curve (see H02 and J03). These 
short-term variations make the 
fitted value of $\alpha_1$ dependent on the sampling of the light curve. 
This is reflected in the formally inconsistent value 
$\alpha_1 = -0.83 \pm 0.04$ obtained by H02, who had a very 
different sampling of the early light 
curve. The true uncertainty in $\alpha_1$ is 
therefore more likely around 0.10. The rapid variations present in
the early light curve are explored in J03. 
\par
We note that there is an indication of a bump in the 
light curve around 26 days after the burst, corresponding to 
$\sim$8 days in the restframe. This corresponds to an extra 
contribution to the flux of the expected smooth power-law decay 
of $\textrm{CL} = 28.92 \pm 0.45$. It is unclear whether a SN could 
produce such a bump since Kirshner et al. (\cite{kir}) find 
very little flux below a restframe wavelength of 
$\sim$2900~\AA\ for a well studied type Ia SN.
%-------------------The X-ray light curve---------------------------------
\section{The X-ray light curve}
\label{xray.sec}
The observations of GRB~011211 by the orbiting XMM-Newton X-ray
telescope started on 2001 Dec 12.30, $\sim$12~hours after the
GRB. We have analyzed
data from the European Photon Imaging Camera (EPIC), using both the
MOS and pn instruments. The total 
observation had a dur\-ation of 29.8~ks for the EPIC-pn 
detector, providing most of the X-ray photons from the afterglow. 
\par
%-------------------FIGURE-----------------------------------------
%----------------------END FIGURE---------------------------------
The data were extracted from the Science Operations 
homepage.\footnote{\texttt{\tiny{http://xmm.vilspa.esa.es/external/xmm\_news/items/grb011211/index.shtml}}}
In order to produce a clean light curve the data were screened
according to the following criteria: (i) only good X-ray events with
single and double pixel events were included, (ii) only events 
in the well calibrated energy range 0.5--10 keV were included, and (iii)
finally a few short periods (much shorter than the time scale of the
variations seen in the light curve) affected by high background were
excluded.
Applying these criteria a light curve was extracted for each detector
from a circular region with a radius of $40\arcsec$ centered on the X-ray
afterglow. Background light curves 
for each detector were produced from two circular
regions with a radius of $40\arcsec$ and centered at the same 
distance from the
nearest chip gap as the X-ray afterglow. For each of the three detectors
the two background light curves were identical within the uncertainties.
The EPIC-pn afterglow flux was corrected for flux falling 
on a chip gap in 
the initial 1.22~hours of the observation and during the subsequent
770~s repointing. Likewise, the background subtracted light curves from the
three detectors showed consistent features and a merged, background
subtracted light curve was produced.
\par
A power-law was fit through the X-ray light curve. The overall
flux decays during the observation with 
a decay index of \mbox{$\alpha_{\mathrm{X}} = -1.96 \pm 0.16$}, 
consistent with the value found by R02. 
As detailed in J03, the first two hours in the X-ray light
curve are most likely affected either by energy variations 
within the expanding
jet, or by emission line features (R02). The X-ray decay slope
increases to \mbox{$\alpha_{\mathrm{X}} = -1.62 \pm 0.36$} if the
initial two hours are omitted from the fit.
\par
In order to compare the optical SED  
(see Sect.~\ref{sed.sec}) to the X-ray spectrum, we extracted 
a 10~ks spectrum from the EPIC-pn detector centered at Dec 12.37.
EPIC-pn data were extracted as described above, and a power-law with
absorption fixed at the Galactic value was fit to the spectrum.
This model is a good fit to the spectrum and absorption in excess of
the Galactic value is not required. The best fit spectral index is
$\beta_\mathrm{X} = -1.21^{+0.10}_{-0.15}$, and the X-ray spectrum 
above 1~keV (which is unaffected by absorption) is shown 
in the inset of Fig.~\ref{sed.fig}.
%-----------------------SED-------------------------------------
\section{Spectral energy distribution of the afterglow}
\label{sed.sec}
%-------------------FIGURE-----------------------------------------
   \begin{figure}
   \centering
   \resizebox{\hsize}{!}{\includegraphics{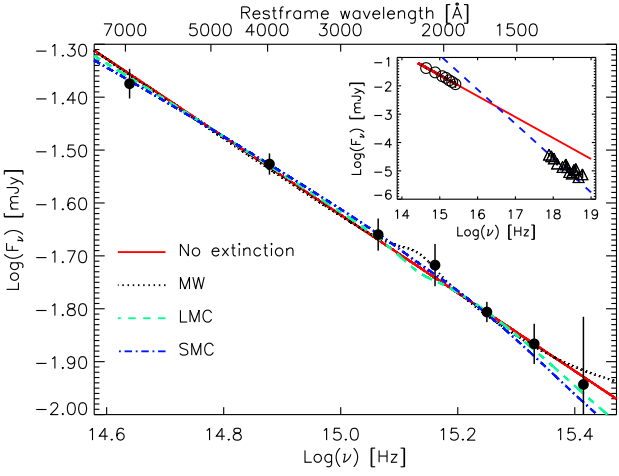}}
      \caption{The spectral energy distribution (SED) 
      of the GRB~011211 afterglow. The data points are based on the 
      $UBVRIJK$-band observations carried out at the NOT, Danish 
      1.54-m and the NTT around Dec 12.3681 
      (see Table~\ref{phot.tab} and J03). The abscissa displays the 
      frequency/wavelength in the GRB restframe. The SED is well 
      fit by an underlying $\beta = -0.56$
      power-law with an SMC extinction law with $A_V = 0.08$. The
      inset shows the $UBVRIJK$-band observations (open rings) along
      with the X-ray spectrum (open triangles) on Dec 12.37. The
      solid line is a pure power-law fit to the optical while the dashed
      line is a pure power-law fit to the X-ray data points.}
      \label{sed.fig}
   \end{figure}
%----------------------END FIGURE---------------------------------
In order to estimate the SED of the 
afterglow and the value of the spectral slope we used the $UJK$ data 
in Table~\ref{phot.tab} along with $BVRI$ data from J03. 
We interpolated the magnitudes to a common epoch, Dec 12.3681 (0.5699 
days after the burst), taking into account the short-term
variations present in the optical light curve (see J03).
We note
that the flux from the host galaxy (estimated in Sec.~\ref{hst.sec}) 
has not been subtracted. However,
the host is faint enough that it will contribute only $\approx$2$\%$ of
the flux at the epoch we are exploring. Therefore, we believe that the
flux from the host galaxy does not significantly affect our results. 
The SED was constructed as explained in Fynbo et
al. (\cite{johan2}). The result is shown in 
Fig.~\ref{sed.fig}, where we have
corrected the data for Galactic reddening using the reddening maps of
Schlegel, Finkbeiner \& Davis (\cite{schlegel}). 
The fact that the SED is similar to those of 
other afterglows at similar redshifts (Jensen et al. 
\cite{brian}; Fynbo et al. \cite{johan2}; 
Holland et al. \cite{holli}) strengthens the validity of our
interpolation approach.
%---------------------------------TABLE----------------------------
\begin{table}
\caption[]{Fits to the GRB~011211 afterglow SED. We note that 
$\chi^2_{\mathrm{dof}}= \chi^2/
\mathrm{degree~of~freedom}$, is the reduced $\chi^2$ of the fit. The
low value of $\chi^2_{\mathrm{dof}}$ in all fits may indicate that we
have overestimated the errors.}
     \label{sed.tab}
\centering
%\tiny
%\scriptsize 
\setlength{\arrayrulewidth}{0.8pt}   % Default is 0.4pt
\begin{tabular}{lccr}
\hline
\hline
\vspace{-2 mm} \\
Fitting function & $\chi^2_{\mathrm{dof}}$ & $\beta$ & 
$A_V$ \hspace{0.4cm} \\
\vspace{-2 mm} \\
\hline
\vspace{-2 mm} \\
No extinction & 0.38 & $-0.74 \pm 0.05$ & 0 \hspace{0.57cm} \\
MW & 0.38 & $-0.82 \pm 0.15$ & $-0.07 \pm 0.12$ \\
LMC & 0.40 & $-0.60 \pm 0.28$ & $0.08 \pm 0.16$ \\
SMC & 0.25 & $-0.56 \pm 0.19$ & $0.08 \pm 0.08$ \\
%$A_V \propto \lambda^{-1}$ & 0.17 & $-0.50 \pm 0.22$ & $0.14 \pm
%0.12$ \\
\vspace{-2 mm} \\
\hline
\end{tabular}
\end{table}
%---------------------------END TABLE--------------------------------
\par
In order to quantify the effects of extinction, we have in 
Fig.~\ref{sed.fig} fit the function $F_{\nu} 
\propto  \nu^{-\beta} \times 10^{-0.4 A_{\nu}}$ to the SED, where 
$A_{\nu}$ is the extinction in magnitudes at frequency $\nu$. We have
considered the three extinction laws ($A_{\nu}$ as a function of
$\nu$) given by Pei (\cite{pei}), i.e. for the Milky Way (MW), Large
Magellanic Cloud (LMC), and Small Magellanic Cloud (SMC). The latter
two are particularly interesting since they have lower abundances of
heavy elements and dust than the MW. In this respect, they may
resemble galaxies at high redshifts, which are presumably in early
stages of the chemical enrichment. 
In these three cases the dependence of the extinction with
$\nu$ has been parameterized in terms of (restframe) $A_V$. Thus, the
fits allow us to determine $\beta$ and $A_V$ simultaneously. 
Finally, we also considered the 
no-extinction case where $F_{\nu} \propto \nu^{-\beta}$.
\par
The parameters of the fits are shown in Table~\ref{sed.tab}. For the
no-extinction case we find a value of $\beta$ consistent with the one
found by H02 \mbox{($\beta=-0.61 \pm 0.15$)}. On the other hand, the 
best fit was achieved for the SMC extinction law, where we
derive a modest extinction of $A_V = 0.08 \pm 0.08$ (restframe $V$)
and a spectral index of $\beta = -0.56 \pm 0.19$.
\par
%---------------------------------TABLE----------------------------
\begin{table*}[t]
\caption[]{Calculation of the closure
relation, $|\alpha_1| + b |\beta| + c$, for two afterglow
models. The closure relation will have a value of zero for a
successful model. The ISM and wind models are for isotropic
expansion in a homogeneous and wind-stratified medium,
respectively. The electron energy power-law index, $p$, is written
as a function of the observed $\beta$, while $\Delta \alpha$ is
calculated from $p$.}
     \label{samanburdur.tab}
\centering
%\tiny
%\scriptsize 
\setlength{\arrayrulewidth}{0.8pt}   % Default is 0.4pt
\begin{tabular}{cccrrc}
\hline
\hline
\vspace{-2 mm} \\
Model & $\nu_{\mathrm{c}}$ & $(b,c)$ & Closure \hspace*{1mm} & 
$p$ \hspace*{11mm}& $\Delta \alpha$ \\
\vspace{-2 mm} \\
\hline
\vspace{-2 mm} \\
ISM & $\nu_{\mathrm{c}} > \nu_{\mathrm{o}}$ & $(-3/2,0)$ & 
$0.11 \pm 0.29$ & $1 - 2 \beta = 2.12 \pm 0.38$ & $(p+3)/4 = 1.28 \pm
0.10$ \\
ISM & $\nu_{\mathrm{c}} < \nu_{\mathrm{o}}$ & $(-3/2,1/2)$ & 
$0.61 \pm 0.29$ & $-2 \beta = 1.12 \pm 0.38$ & $(p+2)/4 = 0.78 \pm 0.10$ \\
Wind & $\nu_{\mathrm{c}} > \nu_{\mathrm{o}}$ & $(-3/2,-1/2)$ & 
$-0.39 \pm 0.29$ & $1 - 2 \beta = 2.12 \pm 0.38$ & $(p+1)/4 = 0.78 \pm 0.10$ \\
Wind & $\nu_{\mathrm{c}} < \nu_{\mathrm{o}}$ & $(-3/2,1/2)$ & 
$0.61 \pm 0.29$ & $-2 \beta = 1.12 \pm 0.38$ & $(p+2)/4 = 0.78 \pm 0.10$ \\
%Jet & $\nu_{\mathrm{c}} > \nu_{\mathrm{o}}$ & $(-2,-1)$ & 
%$-1.45 \pm 0.10$ & $1.66 \pm 0.06$ \\
%Jet & $\nu_{\mathrm{c}} < \nu_{\mathrm{o}}$ & $(-2,0)$ & 
%$-0.45 \pm 0.10$ & $1.66 \pm 0.06$ \\
\vspace{-2 mm} \\
\hline
\end{tabular}
\end{table*}
%---------------------------END TABLE--------------------------------
For the redshift of the GRB~011211 the interstellar extinction bump at
2175~\AA\ is shifted close to the $R$-band. This extinction feature is
very prominent for the MW, moderate in the LMC, and almost nonexistent
for the SMC extinction curve. Our data sampling makes it difficult to 
infer about the presence of a redshifted 2175~\AA\ absorption bump in the
$R$-band. However, 
the best MW fit implies an unphysical negative extinction (see 
Table~\ref{sed.tab}). This result is further strengthened by the fact
that the SMC is consistently a much better fit than the MW for GRBs where
these fits have been applied 
(GRB~000301C: Jensen et al. 
\cite{brian}; Rhoads \& Fruchter \cite{vegir2}; GRB~000926: Fynbo et al. 
\cite{johan2}; GRB~021004: Holland et al. \cite{holli}).
\par
Our best-fit extinction model is consistent with a zero extinction.
This result is strengthened by the fact that
the X-ray spectrum implies no significant absorption in 
the host galaxy (Pedersen et al. \cite{kp}). Furthermore, 
the $A_{V}$ value estimated from the 
$UBVRIJK$-band SED is (although close to the actual value) an upper 
limit of $A_{V}$. This is because the unextincted optical/NIR SED 
segment is not an idealized pure power-law (Sari, Piran \& Narayan 
\cite{sari98}), as
there could be some shallow intrinsic curvature due to the proximity 
of $\nu_{\mathrm{c}}$ to the $U$-band (Granot \& Sari \cite{gs}). 
%In addition, the host was
%not detected in a sub-mm search (Berger et al. \cite{edo2}),
%lending support to our observation of no significant extinction.
In conclusion, the data
support a scenario of a host with a low intrinsic extinction and
which is in the early stages of chemical enrichment.
%------------Comparison with afterglow models----------------------
\section{Comparison with afterglow models}
\label{models.sec}
The parameters $\alpha_1$, $\alpha_2$ and $t_{\mathrm{b}}$, and
$\beta$ can be used to investigate the physical mechanisms responsible 
for the break and the nature of the ambient medium in which the burst 
occurred. Breaks have been observed in many GRB light curves to date. 
They have been interpreted as evidence that the outflows from the bursts 
are collimated with opening angles of approximately 
$5^{\circ}$--$10^{\circ}$ (e.g. Rhoads \cite{vegir}; 
Sari, Piran \& Halpern \cite{sari};
Castro-Tirado et al. \cite{castro}; Holland et al. \cite{holland3}).
If GRBs are collimated outflows, the total energy requirement drops by
a factor of between roughly 100 and 1000, providing a solution to the
so-called ``energy crisis'' of GRBs.
\par
The decay and spectral slopes depend
on the electron energy distribution index $p$. This led 
Price et al. (\cite{price}) and Berger et al. (\cite{edo}) to
introduce the so-called closure relation, in order to distinguish
between various afterglow models. Its exact representation depends on
the definition of $\alpha_1$, $\alpha_2$ and $\beta$. In our notation
$|\alpha_1| + b |\beta| + c = 0$, where the values of $b$ and
$c$ depend on the location of the cooling frequency,
$\nu_{\mathrm{c}}$, relative to the optical/NIR bands, 
$\nu_{\mathrm{o}}$, at the epoch of the observations. We 
use the spectral index found in Sect.~\ref{sed.sec}, $\beta=-0.56 \pm
0.19$, which has been corrected for host extinction. 
Table~\ref{samanburdur.tab} lists two models used in the closure
relation in order to explore the GRB environment before the observed
break in the light curve: (i) expansion into a homogeneous
medium, and (ii) expansion into a wind-fed medium.
\par
Models with $\nu_{\mathrm{c}} < \nu_{\mathrm{o}}$ are disfavoured by the
data. This result is further strengthened by the fact that 
$\alpha_{\mathrm{X}}$ (the X-ray decay index, see Sect.~\ref{xray.sec}) is 
steeper than $\alpha_1$. This implies that 
there is a spectral break between the optical and X rays. 
In addition, our observations imply that the difference
between the low- and high-energy slopes is $\beta - 
\beta_{\mathrm{X}} = 0.65^{+0.21}_{-0.24}$, consistent with 
the prediction
of $0.5$ in the standard synchrotron model (in the slow cooling
regime). This result is displayed in the inset of 
Fig.~\ref{sed.fig}. At the time of the measurements the cooling
frequency is observed to be positioned close to 
$\sim$10$^{16}$~Hz in the observer frame.
\par
Only the ISM ($\nu_{\mathrm{c}} > \nu_{\mathrm{o}}$) model 
produces a 
closure consistent with zero. In Table~\ref{samanburdur.tab} 
we also estimate $\Delta \alpha =
\alpha_1 - \alpha_2$ for the afterglow models. The observed value
of $\Delta \alpha = 1.16 \pm 0.07$ clearly favours the ISM 
($\nu_{\mathrm{c}} > \nu_{\mathrm{o}}$) model. We note that each
spectral or temporal power-law index relates to a certain value of
$p$, and the correct model should result in a similar $p$ for all
indices. In our favoured model we get $p (\alpha_1) = 2.27 \pm 0.03$, 
$p (\alpha_2) = 2.11 \pm 0.07$, $p (\beta) = 2.12 \pm 0.38$, 
$p (\beta_{\mathrm{X}}) = 2.4^{+0.2}_{-0.3}$, and 
$p (\alpha_{\mathrm{X}}) = 2.8 \pm 0.5$. For the X-ray
temporal slope we have used $\alpha_{\mathrm{X}} = -1.62 \pm 0.36$ in
order to avoid the influence of the short-term
variations (see Sect.~\ref{xray.sec}).
\par
We note that $\alpha_{\mathrm{X}} = -1.62 \pm 0.36$ 
is inconsistent with $\alpha_2$,
which makes a chromatic break due to the cooling frequency moving through
the optical band (Covino et al. \cite{covino}) 
inconsistent with the data. Hence, the break in the optical light 
curve is indeed most likely due to a collimated outflow geometry.
\par
The opening angle, $\theta_0$, at the time of the break can be estimated 
using equation~2 in Frail et al. (\cite{frail2}). Using the same 
assumptions as H02 
($\eta_{\gamma} = 0.2$ and $n = 0.1$~cm$^{-3}$) leads to
$\theta_0 = 3.4^{\circ} \pm 0.1^{\circ}$ for 
$t_{\mathrm{b}} = 1.56 \pm 0.02$~days. From this we estimate 
that the \mbox{total} beamed
energy in gamma-rays for GRB~011211 was 
\mbox{$E_{\gamma}\approx 1.2 \times 10^{50}$~erg}, after correcting 
for the beam geometry. This energy is in the
low end of the distribution of the ``standard'' total beamed energy in
gamma-rays centred on $1.3 \times 10^{51}$~erg (Bloom, 
Frail \& Kulkarni \cite{energy}). As pointed out by 
Bloom, Frail \& Kulkarni (\cite{energy}), modeling yields estimates in
the range $0.1~\textrm{cm}^{-3} \lesssim n \lesssim 
30~\textrm{cm}^{-3}$, with little support for extremes of either high
or low density. Assuming $n = 30$~cm$^{-3}$ still gives a relatively
low energy, \mbox{$E_{\gamma}\approx 5 \times 10^{50}$~erg} 
(with $\theta_0 \approx 6.9^{\circ}$), compared
to the median energy of $1.3 \times 10^{51}$~erg.
%---------------------------DISCUSSION---------------------------------
\section{Discussion}
\label{dis.sec}
We have detected a break in the optical light curve of GRB~011211. Our 
observations imply 
($\alpha_1$, $\alpha_2$) = ($-0.95 \pm 0.02$, $-2.11 \pm 0.07$), with
a break time of $t_{\mathrm{b}} = 1.56 \pm 0.02$~days. The SED at
December 12.37 reveals an SMC-like extinction in the
host galaxy at the modest level of $A_V = 0.08 \pm 0.08$, with 
$\beta = -0.56 \pm 0.19$. These properties of the light curve 
could be explained by a jet 
expanding into an ambient medium that has a constant mean
density. We estimate that 
$\theta_0 \approx 3.4^{\circ}$--$6.9^{\circ}$ at the time of 
the break, which reduces the energy released by the GRB by a 
factor of $\sim$100--600 to
($1.2$--$5) \times 10^{50}$~erg.
\par
Using HST/STIS data we estimate the host magnitude to be 
$R = 24.95 \pm 0.11$, a representative value for host galaxies which
typically measure $R=24$--26 (see e.g. Bloom, Kulkarni \& Djorgovski 
\cite{bkd}). 
\par
Finally, we use the various relationships between the light curve 
decay indices ($\alpha_1$, $\alpha_2$ and 
$\alpha_{\mathrm{X}}$) and the spectral
indices ($\beta$ and $\beta_{\mathrm{X}}$) with the electron energy
index ($p$) to calculate five independent values of the latter. We get
an average value of $\sim$2.3, consistent with other bursts that seem
to be adequately fit with models where $p \approx 2.3$--2.5 (van 
Paradijs, Kouveliotou \& Wijers \cite{chryssa}).
%-------------------------acknowledgements------------------------------
\begin{acknowledgements}
     It is a pleasure to thank Avi Loeb, Gunnlaugur Bj\"ornsson, 
     Edo Berger, and Stephen Holland for 
     helpful comments and suggestions. The authors are grateful 
     to the referee, Dr. E. Nakar, for a very constructive report that
     helped us improve the paper on several important points. 
     The data presented here have been taken using ALFOSC, which is
     owned by the
     Instituto de Astrofisica de Andalucia (IAA) and operated at the 
     Nordic Optical Telescope under agreement between IAA and the 
     NBIfAFG of the Astronomical Observatory of Copenhagen.
     Additionally, the availability of the GRB Coordinates Network
     (GCN) and BACODINE services, maintained by Scott Barthelmy, is 
     greatly acknowledged. PJ would like to acknowledge support 
     from NorFA, The Icelandic Research Fund for Graduate
     Students, and a Special Grant from the Icelandic Research Council. 
     JPUF and KP acknowledge support from the 
     Carlsberg foundation. JG acknowledges the receipt of a Marie
     Curie Research Grant from the European Commission. MIA
     acknowledges the Astrophysics group of the Physics dept. of
     University of Oulu for support of his work. This work was 
     supported by the Danish Natural Science Research Council (SNF). 
     The authors acknowledge benefits from collaboration within 
     the EU FP5 Research Training Network "Gamma-Ray Bursts:  An 
     Enigma and a Tool".
\end{acknowledgements}

\end{document}